\documentclass[5p,times,number]{elsarticle}
\pdfoutput=1
   
\usepackage{graphicx,miller,bm,booktabs,float,nicefrac,url,breakurl,color,microtype}
\usepackage{siunitx}[=v2]
\usepackage[version=4]{mhchem}
\usepackage{hyperref}
\usepackage[export]{adjustbox}
\usepackage{upgreek}

\hypersetup{breaklinks,hidelinks}

\DeclareMathAccent{\wtilde}{\mathord}{largesymbols}{"65}

\journal{Scripta Materialia}

\begin{document}

\begin{frontmatter}

\title{The influence of dislocations on R-phase transformations in a NiTi shape memory alloy}

\author[a]{Himanshu Vashishtha}
\author[a]{David M. Collins}
\ead{dmc51@cam.ac.uk}

\affiliation[a]{organization={Department of Materials Science and Metallurgy, University of Cambridge},
        addressline={27 Charles Babbage Rd}, 
        city={Cambridge},
        postcode={CB3 0FS}, 
        country={United Kingdom}}

\begin{abstract}

The ability to control the stress-induced phase transformation of the shape memory alloy, NiTi, is an important technological challenge that must be understood for their wide application in devices that can exploit their reversible strain properties. This study elucidates the direct relationship between dislocation density and the \textit{R}-phase transformation, including its formation temperature from interrupted annealing of rolled NiTi samples. Deformation is shown to determine the enthalpy change required for the B2$\rightarrow$\textit{R}-phase transformation, with associated transformation temperatures being modifiable via dislocation density and recovery processes. Recovery is shown to be rapid, highly heterogeneous and sensitive to crystal orientation. Grains with a $\langle100\rangle$ direction close to the macroscopic rolling direction recover more rapidly than $\langle110\rangle$ and $\langle111\rangle$ orientated grains. Considered to be governed by processing induced residual stresses and resultant crystallographic dependent annihilation/slip pathways, there are opportunities to tune B2$\rightarrow$\textit{R}-phase transformation on either a grain-averaged or an orientation dependant per-grain basis. 
\end{abstract}


\end{frontmatter}

Near equiatomic NiTi shape memory alloys uniquely undergo a stress-induced structurally displacive martensitic transformation, resulting in the macroscopic shape change of the material. When deformed to large strains ($\sim5\%$), these alloys possess a remarkable ability to return to their original crystal structure when heated, called the shape memory effect \cite{YangC, ShariatBS}. During deformation or cooling, the primary B2 phase (austenite, \textit{A}) can transform into twin variants of a monoclinic B19' (martensite, \textit{M}); this may occur directly (i.e. B2$\rightarrow$B19') or via the formation of an intermediary rhombohedrally distorted martensite structure, called the \textit{R}-phase (i.e. B2$\rightarrow$\textit{R}$\rightarrow$B19') \cite{OtsukaK, FengB}. Promoting the \textit{R}-phase formation and understanding its mechanistic role to mediate deformation is attractive\cite{FengB}; however, NiTi has limited application whilst it suffers from a diminishing memory effect with repeated thermal or load cycling. This is postulated to be controlled by the accumulation of dislocations \cite{AkamineH, JonesN},  controlled by grain neighbour constraint, crystallographic orientation related plastic anisotropy, and martensite variant selection \cite{LaplancheG}. Such factors are critical as transformation behaviour is dominated by local stresses \cite{Sedmak}.

\begin{figure}[h!]
\centering
    \includegraphics[width=90mm]{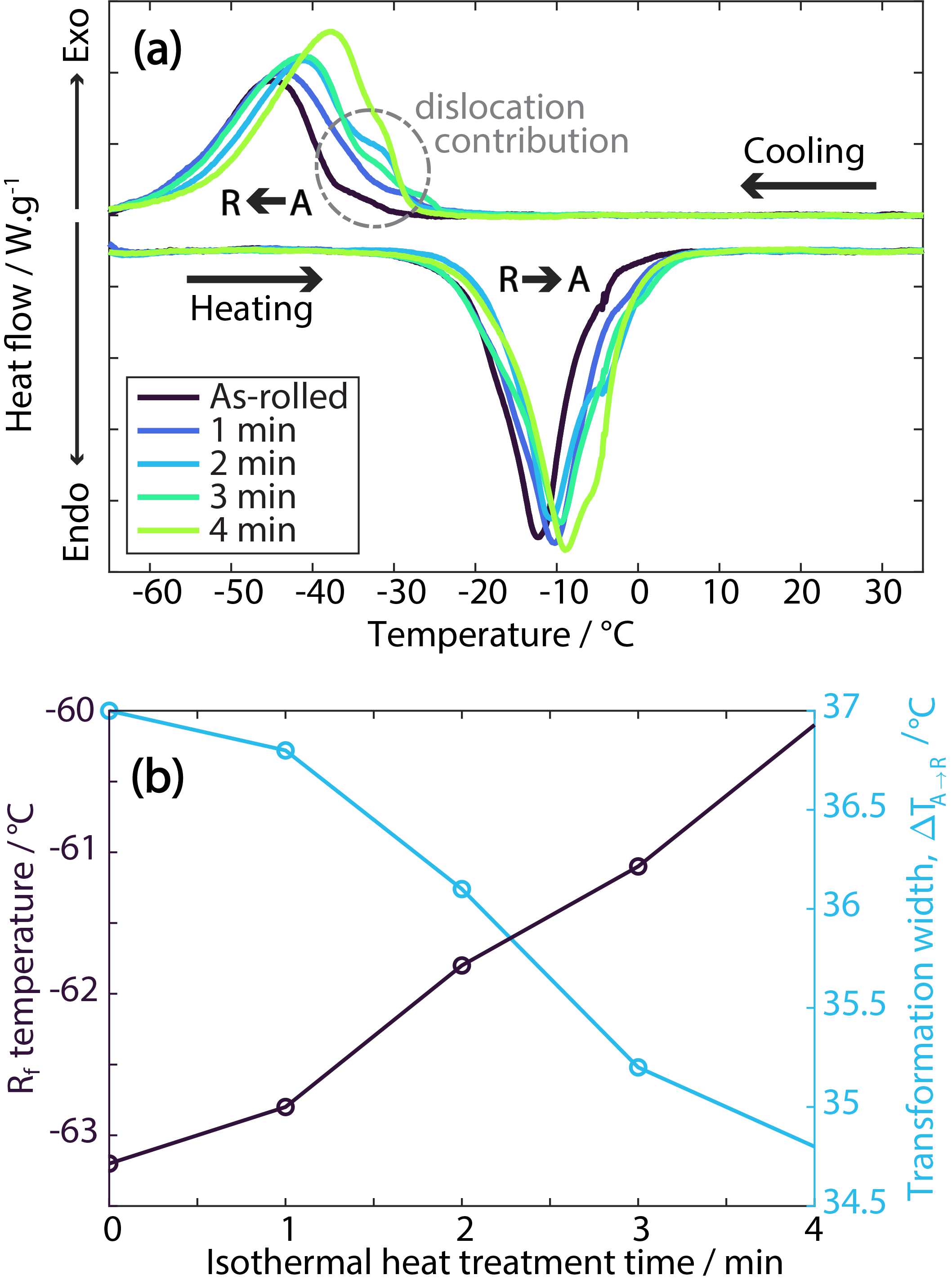}
    \caption{The B2$\leftrightarrow$\textit{R} transformation is shown in (a) via DSC plots during heating and cooling for the as-rolled and isothermally heat treated specimen for 1, 2, 3 and 4 mins. The \textit{R}-phase phase transformation temperatures were measured using the tangent intersection method; this and the transformation width is shown in (b).}
    \label{fig:DSC}
\end{figure}

An additional challenge for the exploitation of near-equiatomic NiTi is to understand and control cyclic changes to the transformation temperature, reported to be $\sim$20\,K after a few hundred loading cycles \cite{AkamineH, PattabiM}. With a lower energy barrier for its formation, the \textit{R}-phase is hypothesised to experience smaller transformation temperature changes and, critically, has less memory-loss associated damage compared to the direct B2$\leftrightarrow$B19' transformation. The \textit{R}-phase may also be considered attractive due to its low heat capacity, giving a fast response to temperature variations\cite{LemkeJM}, a low thermal hysteresis \cite{WangXB} and high functional stability during thermomechanical cycling \cite{WangX}.  Furthermore, property degradation could be limited by the crystallographic compatibility afforded by the \textit{R}-phase to the B2 and B19' phases \cite{GuH}. Both bulk chemistry and thermomechanical treatments can promote the \textit{R}-phase occurrence \cite{WangXB}, and specifically stabilised by tailoring i) the chemical composition (replacing 1-2\% of Ni with Fe or Al, albeit with a change to the \textit{R}-phase and \textit{M} transformation temperatures, ii) the coherency strains associated with the Ni$_4$Ti$_3$ precipitates, and iii) lattice distortion from dislocations  \cite{ZhuJ, ZhaoY}. Processing related factors that control microstructure and remnant deformation, however, remains seldom explored in context to the \textit{R}-phase formation. In this study, the role of microstructure and plastic deformation that results from sheet rolling on the \textit{R}-phase transformation has been explored. To isolate the role of dislocation alone, isothermal heat treatments were explored that give negligible changes to the microstructure. Differential scanning calorimetry (DSC) and electron backscatter diffraction (EBSD) measurements are used to correlate the enthalpy change and stored elastic strain energy to the dislocation density. 

\begin{figure*}[]
\centering
\vspace{-90pt}
    \includegraphics[width=170mm,center]{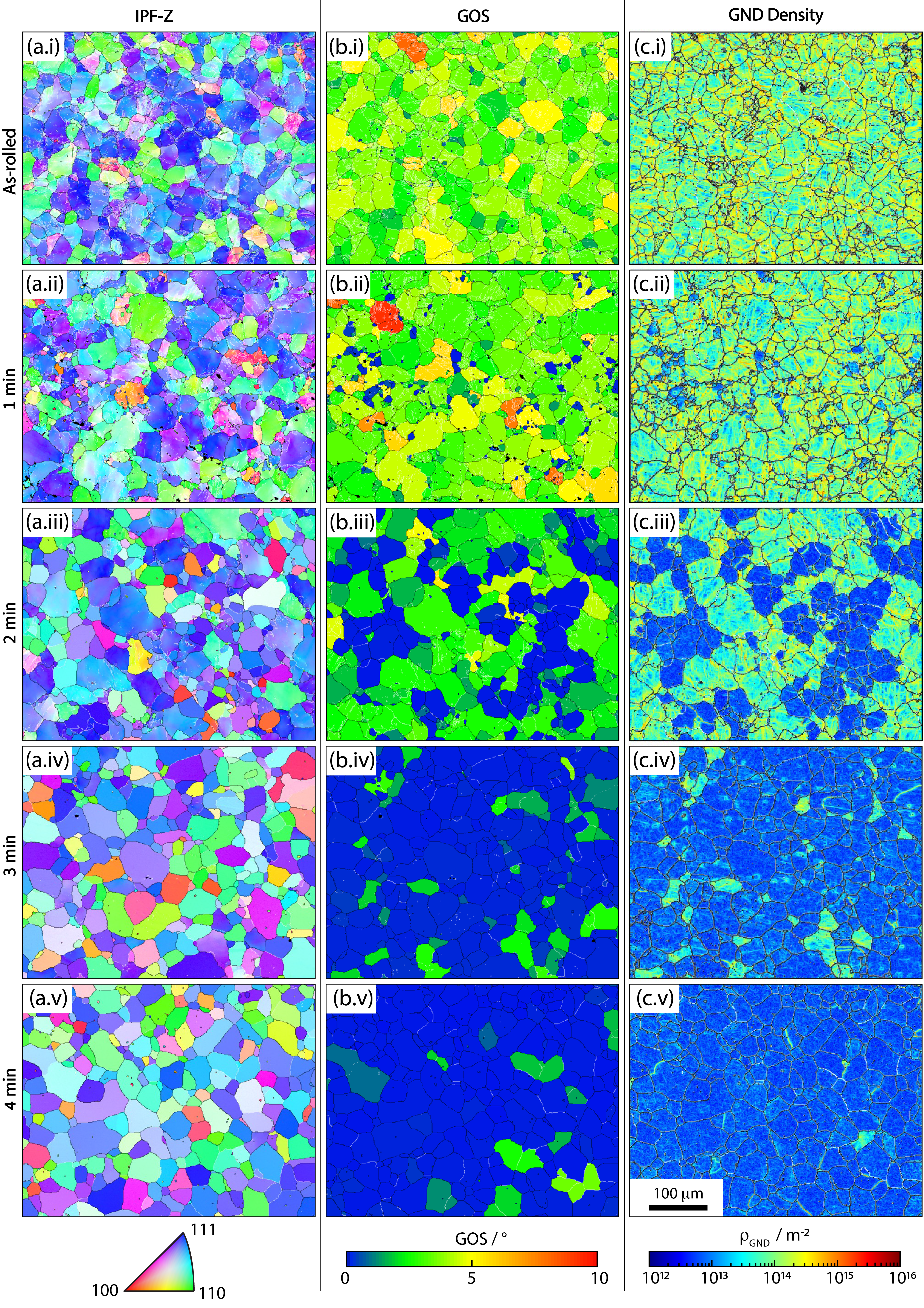}
    \caption{IPF-$Z$ maps shown as (a); GOS maps as (b) and GND maps as (c) for as-rolled, and 1, 2, 3, and 4 minutes isothermally heated specimens. The GOS and GND density maps can be used to infer the progression of recovery during isothermal heating. Here, the rolling direction is left-right, and $Z$ is the normal direction.}
    \label{fig:EBSD}
\end{figure*}

NiTi sheets of 0.5\,mm thickness were produced by \textit{SMA Wires India} to a specific austenitic finish {\textit{A}$_{\rm{f}}$} temperature of $10^\circ$C. The alloy had a nominal composition of 50.8\%Ni-49.2\%Ti, and was supplied in a proprietary rolled condition with a high level of cold work. The material was sectioned into $2$\,mm $\times$ $2$\,mm samples (of mass $\sim 12.9$\,mg) for DSC testing. A NETZSCH DSC 214; samples wereheated \& cooled at 10\,$^\circ$C\,min$^{-1}$ from -65\,$^\circ$C to 120\,$^\circ$C. Using the transformation temperatures inferred from the DSC results, NiTi coupons of $10$\,mm $\times$ $10$\,mm size were subjected to short duration isothermal heat treatments, comprising 1, 2, 3, or 4 minutes at 850\,$^\circ$C. For characterisation, sample surfaces were ground with abrasive media, followed by polishing with a diamond suspension, then electropolished using a Struers LectroPol-5 with a perchloric acid (21\,vol.\%) and acetic acid (79\,vol.\%) electrolyte at 10\,$^\circ$C \cite{VashishthaH}.  Including the as-rolled condition, isothermally heat treated samples were examined via EBSD. For the latter, a ThermoFisher Apreo 2 equipped with an Oxford Instruments Symmetry S3 EBSD detector, operated at 20\,kV, was used. For data collection and diffraction pattern indexing, the software AZtecCrystal was used; EBSD maps were collected from representative areas of the microstructure measuring $500 \times 400$ \,$\upmu$m$^2$ and a step size of $1$\,$\upmu$m. Thereafter, the MTEX package \cite{Mtex} written for Matlab was used for EBSD data processing for deformation quantification. Here, a calculation of the lattice curvature tensor was obtained to estimate the total geometrically necessary dislocation (GND) density, following Pantleon \cite{PantleonW}. The values quoted are the sum of the dislocations on all B2 slip systems, using a Burgers vector, $b = \frac{a}{2}\langle111\rangle$.

The thermal phase transformation behaviour of the as-rolled and isothermally heat treated specimens, measured by DSC, is shown in Fig. \ref{fig:DSC}a. The narrow transformation hysteresis, $<10^\circ$, is characteristic of \textit{R}-phase formation \cite{VashishthaH} and corroborates prior observations that plasticity favours \textit{R}-phase formation \cite{ChrobakD}. During cooling, the transformation curve shifts to higher temperatures with increased heat treatment time. A suppressed feature, labelled as `dislocation contribution', is evident during the $A\rightarrow R$ transformation close to the \textit{R}-phase start temperature, considered here to be related primarily to the degree of plasticity in each sample. When heated, this feature becomes less prominent with increasing recovery time. Moreover, the difference between the \textit{R}-phase start ($R_{s}$) to finish ($R_{f}$) temperature, ($\Delta T_{A\rightarrow R}$), drops with increased isothermal heating time, indicative of a shrinking transformation window. This is plotted in Fig. \ref{fig:DSC}b. The integral of the heat flow function is defined as the enthalpy change, $\Delta H$ \cite{ChekotuJC}, which increases monotonically after heat treatments. However, as the stored elastic strain energy from dislocations promotes the phase transformations \cite{AkamineH}, the total phase transformation energy, $E_{\rm total}$, is defined as:

\begin{equation}
    \Delta E_{\rm total} = f(E_{\Delta H_{A\rightarrow R}}, E_{\rho}) 
\end{equation}

\noindent where $E_{\Delta H}$ is the enthalpy change (exothermic or endothermic), and $E_{\rho}$ is the elastic energy of the dislocations. Here, $E_{\rho}$ drops during recovery as the dislocation density reduces. It is proposed that $E_{\Delta H}$ shall change in the same proportion to meet the required total energy to form the \textit{R}-phase. It is assumed that any change in energy associated with the grain boundary area, is negligible, and that the volume fraction of precipitates is small, such that any localised coherency strain energy, can be ignored.

\begin{figure}[h!]
\centering
    \includegraphics[width=90mm]{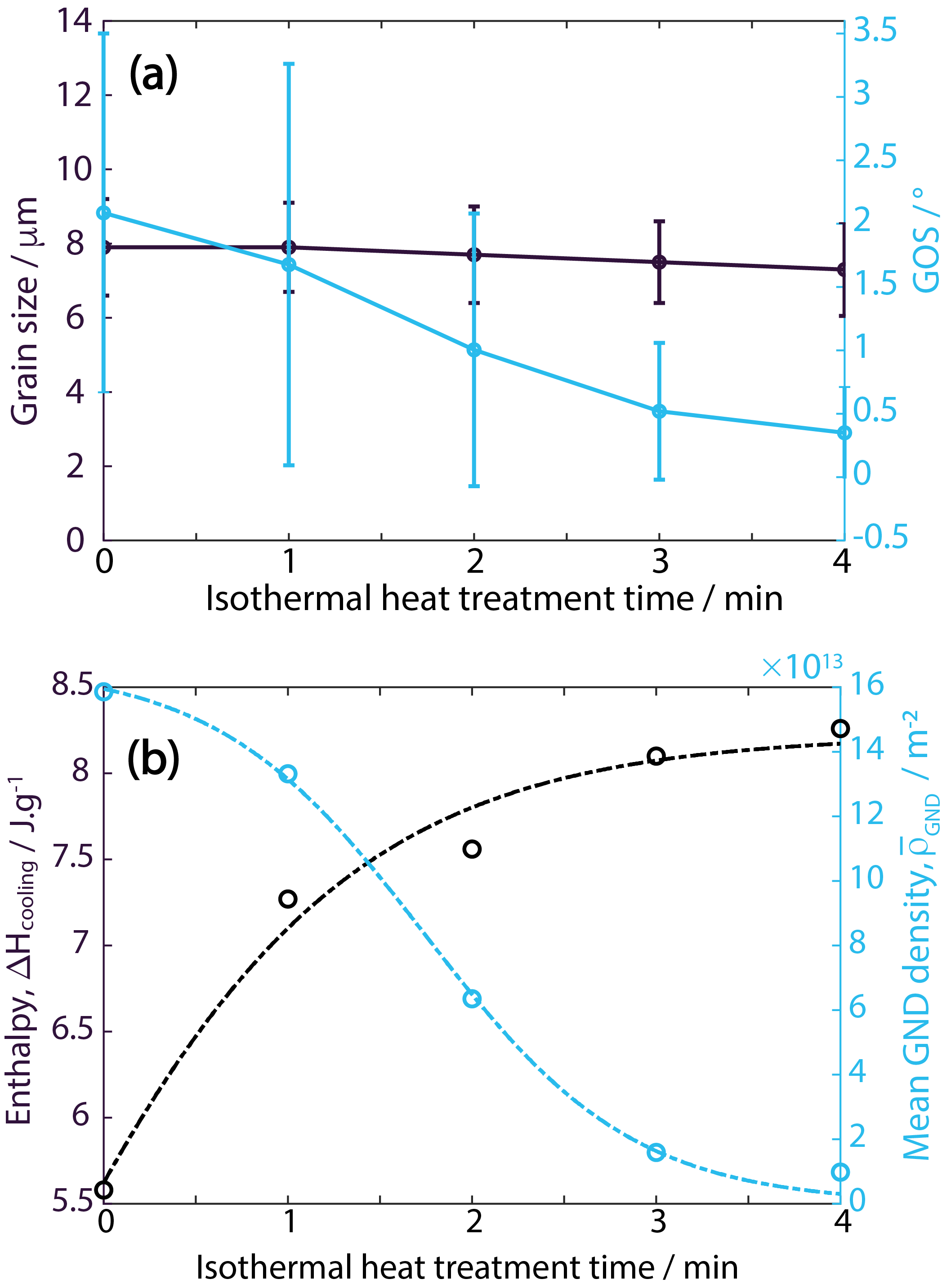}
    \caption{(a) Variation of mean grain size and GOS with isothermal heating time; error bars are the standard deviations against measurements. In (b), enthalpy changes are estimated from DSC measurements, and from EBSD datasets, the mean GND densities are estimated.}
    \label{fig3}
\end{figure}

Inverse pole figures maps from EBSD measurements (IPF-$Z$, normal direction to the sheet) of the as-rolled and isothermally heat treated (1, 2, 3 and 4 mins) samples are shown in Fig. \ref{fig:EBSD}a. Only the B2 cubic austenite phase was present for all specimens; no  Ni$_4$Ti$_3$ precipitates were observed at the resolution employed here. The fraction of low angle grain boundaries (LAGBs, 2-15$^\circ$) decreased with heat treatment time; measured as 73.2 \%, 62.9 \%, 35.9 \%, 13.5 \%, and 11.2 \%, for  the as-rolled, 1, 2, 3, and 4 minute heat treated specimens, respectively. It is well established that the LAGBs are associated with the dislocation structures \cite{YazdandoostF}, indicating that the dislocation density decreases during the heat treatments. Importantly, no significant grain size change was observed for the heat treatments studied, as shown in Fig. \ref{fig3}a, as expected from the short thermal periods. Recovery can be further described using grain orientation spread (GOS) maps, Fig. \ref{fig:EBSD}b, showing  the averaged misorientation angle relative to a grain reference orientation. Such data treatment highlights the recovered grains with low a GOS against those with high remnant plasticity and associated residual strains \cite{SureshKS}, with high GOS. Averaging GOS for each examined condition shows a monotonic decrease, shown in Fig. \ref{fig3}a.

Spatially resolved quantification of the deformation was obtained from an estimate of the GND density, $\rho_{\rm GND}$. In the as-rolled condition, Fig. \ref{fig:EBSD}c.i, the microstructure is highly deformed with all grains experiencing GND densities between $10^{14}$\,m$^{-2}$ and $10^{15}$\,m$^{-2}$. All grains see bands of higher dislocation density, often intersecting high angle grain boundaries where grain boundary constraint is a likely the source for their formation. There are several grains with cellular-type features; structures consistent with dislocation structures that are known to form in cubic systems \cite{WangXZPu}. Once the NiTi has been heated recovery begins, however, proceeds in a highly heterogeneous manner. This is evident after 1 minutes at 850$^\circ$C for small grains in random positions, shown as those with a GND density $<10^{13}$\,m$^{-2}$ (Fig \ref{fig:EBSD}c.ii). Heating for 2 minutes, Fig \ref{fig:EBSD}c.iii, shows an increased fraction of the recovered microstructure, evident in clusters of recovered grains (low GND density regions) surrounded by grains that retain high levels of deformation (high GND density regions). After 3 minutes, Fig \ref{fig:EBSD}c.iv, only a few grains with high levels of deformation remain. Recovery appears complete after 4 minutes, with $\rho_{\rm GND}$ densities close to EBSD detection limits \cite{IspanovityPD}. For completeness, kernel average misorientation (KAM) maps are provided in Fig. S1 of the Supplementary Material.

\begin{figure}[h!]
\centering
    \includegraphics[width=90mm]{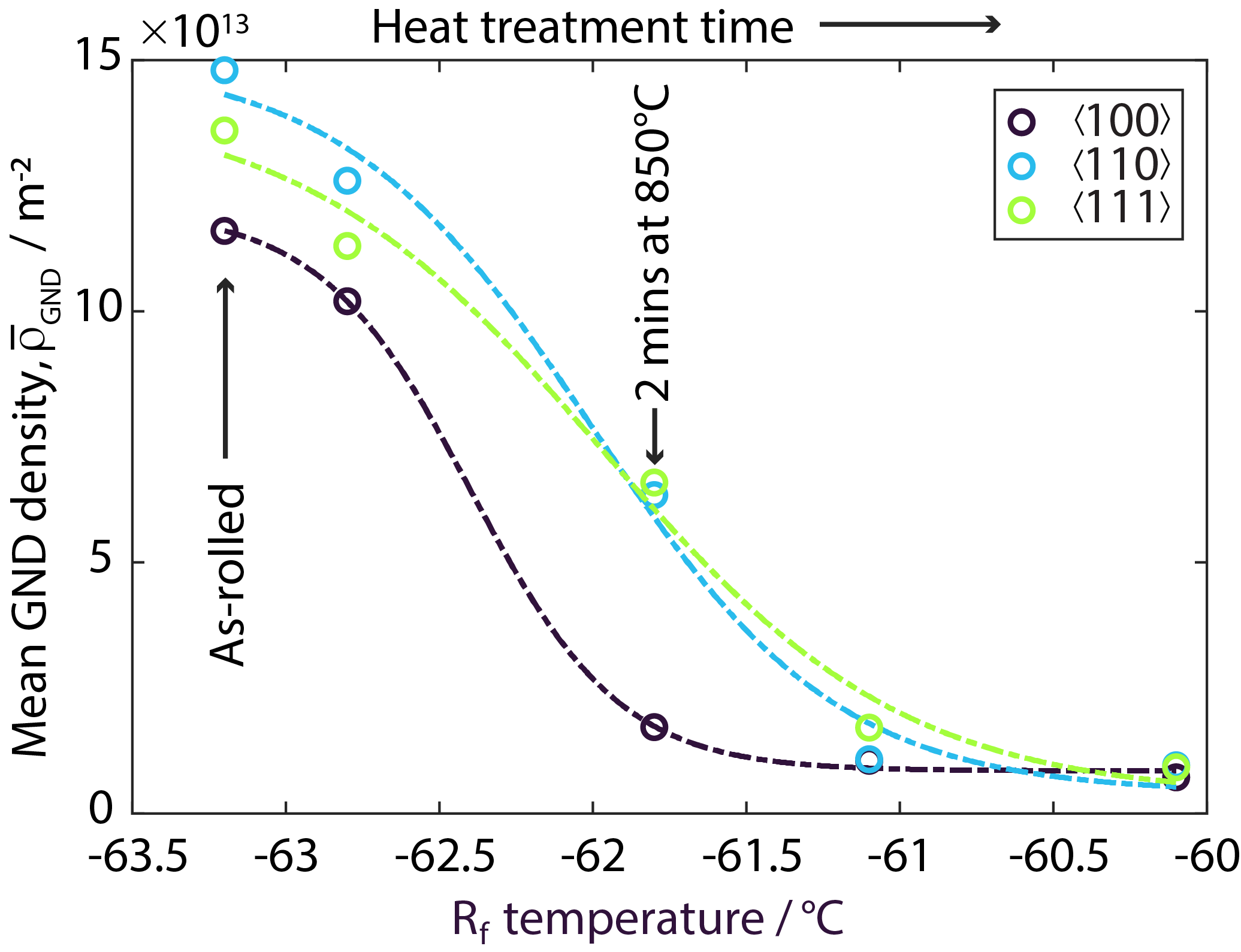}
    \caption{Crystal orientation dependence of the GND density as a function of \textit{R}-phase finish temperature, $R_f$, for grain families with the directions $\langle1 1 1\rangle$, $\langle1 0 1\rangle$ and $\langle0 0 1\rangle$ near to the rolling direction.}
    \label{fig:Orientation}
\end{figure}

With an approximately constant grain size during heating, the transformation enthalpy correlated with $R_f$ and $\Delta T_{A\leftrightarrow R}$ is unambiguously dominated by the dislocation density of each tested condition. Evidently, a high dislocation density in the parent phase creates a barrier for the $A \leftrightarrow M$ transformation, in which its formation must overcome a large lattice distortion ($\sim 10 \%$ \cite{WangXB}). The comparatively small lattice distortion associated with the $R$-phase formation ($\sim1 \%$ \cite{WangXB}) is clearly affected by the presence of plasticity, but its formation is not suppressed. At higher dislocation densities, the $R$-phase becomes more stable than B19', making the B2$\leftrightarrow R$-phase pathway more favoured than the direct B2$\leftrightarrow$B19' transformation. During the recovery process,  $>2$ minutes at $850^\circ$C corresponds to the time required for mobile dislocation segments to move over critical annihilation distance. This period corresponds to the most rapid mean GND density change for each sample, shown in Fig. \ref{fig3}b, with an approximately sigmoidal function describing the recovery process. The interrupted annealing intervals, in context to the stored energy dislocation content, favour the thermal transformations. The energy required (enthalpy change, $\Delta H$) for B2$\rightarrow$\textit{R}  transformation is the lowest for the as-rolled specimen (high $\bar{\rho}_{\rm GND}$ density), as shown in Fig. \ref{fig3}b. Notably, $\Delta H$ increases with isothermal heating time, even though the transformation window, $\Delta T_{A\rightarrow R}$, drops, becoming smaller when the transformation itself is dominated by enthalpy alone. As $\Delta H$ follows the reverse trend of $\bar{\rho}_{\rm GND}$, Fig. \ref{fig3}b, the enthalpy is lower when elastic strain energy from dislocations can facilitate the \textit{R}-phase formation (i.e. $\Delta E_{\rm total}|_{T,t} \approx E_{\Delta H_{A\rightarrow R}} + E_{\rho})$. It is assumed that the measurable dislocation content, $\rho_{\rm GND}$ is indicative of the total dislocation density, $\rho$, with the statistically stored dislocations, $\rho_{\rm SSD}$, comprising sessile dislocation dipoles with limited capability to annihilate during recovery, and so can be neglected.

Curiously, a grain orientation dependence of recovery was observed. This trend is summarised in Fig. \ref{fig:Orientation} where the mean GND density for similarly orientated grains is plotted as a function of $R_f$ temperature (proportional to heat treatment time). Specifically, grains with crystallographic directions, $\langle111\rangle$, $\langle101\rangle$ and $\langle001\rangle$, within $20^\circ$ of the rolling direction of the sheet are grouped. In the as-rolled state, a small but measurable $\bar{\rho}_{\rm GND}$ density dependent on orientation is present; this is due to the limited slip that is permitted in B2, $\langle1 0 0\rangle\{0 0 1\}$ and $\langle1 0 0\rangle\{0 1 1\}$ \cite{Chumlyakov1996102}, whose activation is dependant on the stress tensor imposed by the plane strain rolling and the crystal orientation itself. This slip geometry explains why the $\langle001\rangle$ grains have been shown to require the highest resolved shear stress for slip activation \cite{SaigalA}, which rationalises why $\langle001\rangle$ grains in this study have the lowest $\bar{\rho}_{\rm GND}$. When heating, however, the disparity between the $\langle100\rangle$ and $\langle110\rangle$/$\langle111\rangle$ grains increases significantly, such that the $\langle100\rangle$ grains recover sooner (see 2 minutes at 850$^\circ$C). As $R_f \propto$ time, Fig. \ref{fig:DSC}b, the \textit{R}-phase onset temperature must be controlled by the grains that possess a near-linear $\rho_{\rm GND}-R_f$ relationship. The $\langle100\rangle$ grains recover first, however, the sigmoidal trend in Fig. \ref{fig:Orientation} indicates their limited influence on the phase transformation temperature, which are instead dominated by the $\langle110\rangle$/$\langle111\rangle$ grains that retain deformation for longer. The orientation dependence of recovery corroborates other studies where $\langle111\rangle$ grains show the least recoverability compared to other orientations \cite{SaigalA, GallK}, and orientation dependence during dynamic recovery \cite{KUBIN20092567}, however, these differ from the present work as the external applied load can be used to explain the per-grain behaviour. In the absence of an externally applied load, the recovery must be controlled by the intragranular and backstress-controlled intergranular residual stress state. As the latter is related to the permitted slip systems, it is proposed this creates sufficient residual stress heterogeneity with crystal orientation, that orientation dependent recovery prevails.

This study has isolated a direct relationship between the dislocation density, the favourability of the \textit{R}-phase transformation and its formation temperature. Specifically, the presence of deformation is shown determine the required B2$\rightarrow$\textit{R}-phase enthalpy change, and the associated transformation temperatures can be tuned through dislocation density and recovery. Orientation-dependent recovery further indicates per-grain transformation are also achievable; it is recommended that such behaviour is explored further. For future NiTi applications that exploit the B2$\rightarrow$\textit{R}-phase transformation, the results here could aid (i) understanding the progressive and inhomogeneous annealing response during thermal processing, and (ii) correlate dislocation density accumulation during loading (monotonic or cyclic) to predictions of onset transformation temperature associated with B2$\rightarrow$\textit{R} phase changes.

\section*{Acknowledgements}

The assistance of Amy Newell for DSC measurements at the University of Birmingham is gratefully acknowledged. Funding from a Leverhulme Trust Research Project Grant (RPG-2021-244) is also acknowledged.

\typeout{}
\bibliography{references.bib}
\bibliographystyle{elsarticle-num-names}

\end{document}